\documentstyle[twoside,fleqn,espcrc2]{article}
\newcommand{\AmS}{{\protect\the\textfont2
  A\kern-.167em\lower.5ex\hbox{M}\kern-.125emS}}

\title{Numerical Results for Transport Coefficients  of Quark Gluon
       Plasma with Iwasaki's Improved Action}
        
\author{A. Nakamura,\address{Research Center for Information Science 
       and Education, Hiroshima\\},
T.Saito,\address{Faculty of Education, 
        Yamamagata University,Yamagata\\}
       and
       S.Sakai$^{\rm b}$
       \thanks{Presented by S.Sakai}}

\begin{document}

\begin{abstract}
%The numerical result for the transport coefficients of quark gluon 
%plasma is obtained by lattice simulation of SU(3) pure gauge model.
Numerical results for the transport coefficients of quark gluon 
plasma are calculated by lattice simulation of SU(3) pure gauge model.
The bulk viscosity is consistent with zero. The shear viscosity is
finite and increases with temperature $T$ roughly as $T^{3}$, and around the 
finite temperature transition points, it is slightly smaller than
%the ordinary hadron masses. 
the typical hadron masses. 

\end{abstract}

% typeset front matter (including abstract)
\maketitle

\section{Introduction}
In the phenomenological study of quark gluon plasma(QGP), when its bulk
%properties are concerned, it is usually treated as gas or liquid.
properties are concerned, the system of quarks and gluons is usually treated as gas or liquid.
Then the fundamental parameters of QGP such as transport coefficients,
are very important information. The aim of this work is to calculate them
from the fundamental theory of QCD.\\
\indent
The calculation of transport 
coefficients is formulated in the framework of linear response theory
of Kubo\cite{Zubarev,Hosoya,Karsch,Horsley}.
 They are expressed by the space time integral of retarded
%Green's function of energy momentum tensor at finite
Green's function of energy momentum tensors at finite
temperature. The shear viscosity $\eta$
is expressed as follows.
\begin{eqnarray} \eta = - \int d^{3}x' \int_{-\infty}^{t} dt_{1} 
e^{\epsilon(t_{1}-t)} \hskip 2cm \nonumber \\   
\times \int_{-\infty}^{t_{1}} dt'<T_{12}(\vec{x},t)T_{12}(\vec{x'},t')>_{ret} 
%\number
\end{eqnarray}
%\begin{equation} 
% \eta = - \int d^{3}x' \int_{-\infty}^{t} dt_{1} e^{\epsilon(t_{1}-t)} 
%\int_{-\infty}^{t_{1}} dt'<T_{12}(\vec{x},t)T_{12}(\vec{x'},t')>_{ret} 
%\end{equation}
Similarly the bulk viscosity $\zeta$ and heat 
conductivity $\chi$ are expressed in terms of the retarded Green's function of 
$T_{11}$ and $T_{41}$ components of 
energy momentum tensor.
%In the pure gauge model the energy momentum tensor is written
In the pure gauge model the energy momentum tensors are written
by the the field strength tensor.\\
% as follows, 
%\begin{equation}
%$T_{\mu \nu}=2Tr[ F_{\mu \sigma}F_{\nu \sigma}
%-\delta_{\mu \nu}F_{\rho \sigma}F_{\rho \sigma}/4]$\\
%$\end{equation}
\indent
%which is defined by the plaquette in the lattice gauge theory.\\
\indent
%The calculation of the retarded Green's function at finite temperature
The direct calculation of the retarded Green's function at finite temperature
is very difficult. Then the shortcut is to calculate 
Matsubara Green's function($G_{\beta}$) and   
then by the analytic continuation, we obtain the 
retarded Green's function at finite temperature.
The analytic continuation is carried out by the use of the fact that 
%the spectral function of fourier transform of the both Green's 
the spectral function of Fourier transform of the both Green's 
functions is the same. 
For the spectral function 
we use the following simplest ansatz\cite{Karsch}.
\begin{eqnarray}
%\begin{equation}
\rho(\vec{p}=0,\omega) \hskip 4cm \nonumber \\
 = \frac{A}{\pi}                                         
(\frac{\gamma}{(m-\omega)^2+\gamma^2}-\frac{\gamma}{(m+\omega)^2+\gamma^2}) 
%\end{equation}
\end{eqnarray}
\noindent                                                                      
where $\gamma$ partially represents the effects of the interactions and
is related to the imaginary part of self energy.
Under this ansatz, the transport coefficients are calculated as,           
\begin{equation}
\alpha \times a^3 = 2A\frac{2\gamma m}{(\gamma^2+m^2)^2}
\end{equation}  
%\begin{equation}              
%(\frac{4}{3} \eta+\zeta) \times a^3 = 
%2A\frac{2\gamma m}{(\gamma^2+m^2)^2}_{11}\\
%\end{equation}                
where $\alpha$ means $\eta$, $\frac{4}{3} \eta+\zeta$ and
$\chi \cdot T$.

\indent
We notice that if $m=0$ or $\gamma=0$, transport coefficient                   
becomes zero.                                                                  
In order to determine these parameters, at least three
independent data points in $G_{\beta}$
are required in the temperature direction.                               
  In the pioneering work of Karsch and Wyld\cite{Karsch}, they performed       
lattice QCD calculations on $8^3 \times 4$ lattice and the resolution was      
not enough to determine these three                                            
parameters independently.                                                      

\section{Numerical Results}
 We carry out our simulation on $16^3 \times 8$ lattice. 
We have started the the simulation
from U(1) gauge theory.  It is found that the fluctuation of
$G_{\beta}$ is very large that it need about
$10^6$ data for the determination of $G_{\beta}$ in the deconfined phase.  
Further in the confined phase, the fluctuations become still larger  
and we could not obtain the
Green's functions even with the $\sim 1.5 \times 10^{6}$ data\cite{Sakai}.
Similar results are obtained in the case of SU(2) gauge theory.
%We think that this is because the energy momentum tensor operator
%in the confined phase should
%be written by the hadron(Glueball) field rather than the gluon
%fields.\\
Then in the calculation of SU(3) gauge theory, we carry out our simulation 
only in the deconfined phase.\\
\indent
It is also found that 
the fluctuation of $G_{\beta}$ 
becomes larger as we proceed to $U(1) \rightarrow SU(2) 
\rightarrow SU(3)$.  Then it is a very important problem to 
reduce the fluctuation of $G_{\beta}$. 
We find that by using the Iwasaki's
improved action, the fluctuation is much reduced as shown in Fig.1.
\begin{figure}[htb]
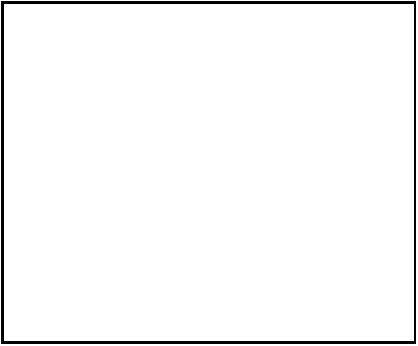

\vspace{9pt} 
\framebox[55mm]{\rule[-21mm]{0mm}{43mm}}
\caption{Time histories of $G_{\beta}$ of $T_{11}$ at $T=2$ with 
improved and standard action}
\label{fig:largenenough}
\end{figure}
%It is seen that the fluctuation is reduced $\sim 1/10$ in 
%the case of the Improved action. 
Then in the following we apply Iwasaki's Improved action for the 
simulation of SU(3) gauge theory.\\
\indent 
From roughly $0.5 \times 10^{6}$ data, we obtain
$G_{\beta}$ for $T_{11}$ and $T_{12}$ .
But they have still rather large errors. 
The fit of $G_{\beta}$ with parameters in    
the spectral function given 
by Eq.(2) is done with SALS. 
The shear and bulk viscosities are calculated by these parameters by 
%Eqs.(3) and the error are estimated by Jackknife method.
Eqs.(3) and the error are estimated by the Jackknife method.
The results for them are shown in Fig.2.
It is found that the bulk viscosity is zero within errors.
\begin{figure}[htb]
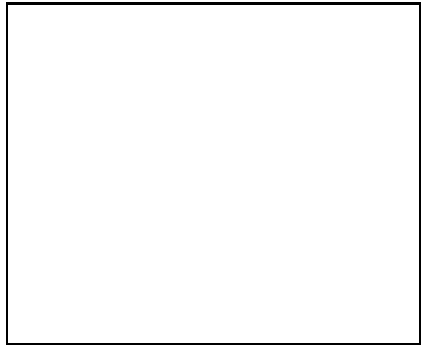

\vspace{9pt}
\framebox[55mm]{\rule[-21mm]{0mm}{43mm}}
\caption{Shear and bulk viscosity $\times a^3$}
\label{fig:largenenough}
\end{figure}
\indent
This result is consistent with the arguments of S. Gavin\cite{Gavin}
that in the pure gauge theory, the bulk viscosity and heat conductivity
should be zero.  In our calculation, $G_{\beta}$ of
$T_{14}$ from which the heat conductivity is calculated, has large 
background and it has been impossible to get signal from it.\\
\indent
In order to know the shear viscosity in the physical unit,
we should determine the lattice spacing $a$ in these $\beta$ values.
For this purpose we have started to determine 
the finite temperature transition points $\beta_c$ at 
$N_{T}=8$. We made a histogram analysis on $16^3 \times 8$ lattice.
The results for the Polyakov susceptibility are shown in Fig.3. 
\begin{figure}[htb]
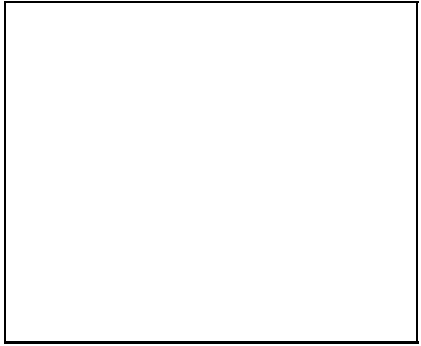

\vspace{9pt}
\framebox[55mm]{\rule[-21mm]{0mm}{43mm}}
\caption{The polyakov susceptibility on $16^3 \times 8$.}
\label{fig:largenenough}
\end{figure} 
\indent
The spacial volume is too small to make a 
precise determination of
$\beta_c$ on our lattice,
but we find that the transition region is 
$2.71 \leq \beta_c \leq 2.73$.  By using the 
finite size scaling formula and the transition temperature 
$T_{c} \sim 276(3)(2)MeV$ determined by Tsukuba 
group \cite{Kaneko} and assuming asymptotic scaling for
$\beta \geq 2.73$ region, the lattice spacing is estimated.\\
\indent
The preliminary results for the shear viscosity in the physical 
unit are shown in Fig.4.
\begin{figure}[htb]
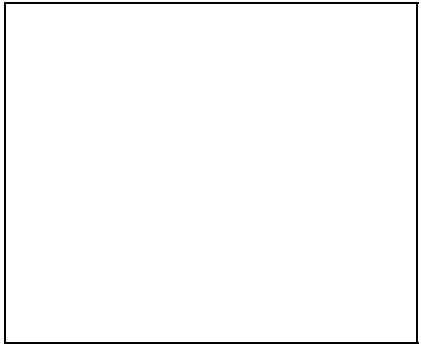

\vspace{9pt}
\framebox[55mm]{\rule[-21mm]{0mm}{43mm}}
\caption{Shear viscosity in the physical unit}
\label{fig:largenenough}
\end{figure} 
It is found that it increases with
temperature.
%It's $T$ dependence is consistent with $T^{3}$, because
The $T$ dependence is consistent with $T^{3}$, because
$\eta \times a^3$ is almost independent of 
$\beta$ as shown in Fig.2 and $a \propto 1/T$ when $N_{T}$ is fixed.  
But clearly more accurate data are necessary.
It is also seen that it seems finite around $T_{c}$, which 
is slightly smaller than the typical hadron massess. 
What is the physical effects on the phenomenology of quark gluon plasma,
when it has shear viscosity with this magnitudes, is a very 
interesting problem.

\section{Conclusions and Further Problem}

We have obtaind the transport coefficients of quark 
%gluon plasma by using Iwasaki's improved action, as shown in Figs.2 and 4.
gluon plasma from the lattice QCD calculation by using Iwasaki's improved action, as shown in Figs.2 and 4.
The shear viscosity increases with temperarure roughly as $T^3$ 

The bulk viscosity is consistent with zero. And the heat conductivity
is very difficult to calculate because of large background for 
$G_{\beta}$ of $T_{14}$.

 Our results depend strongly on the ansatz of spectral function in Eq.(2).
In order to study the functional form for the  spectral function, we have 
%started the simulation with asymmetric lattice.  
started the simulation with anisotropic lattices.  

We could not calculate the $G_{\beta}$ in the confined
phase\cite{Sakai}. We think that this is because the energy momentum 
%tensor operator should be written
tensor operators should be written
by the hadron fields, rather than the gluon fields.  In order to certify
this hypothsis, we have started the calculation of gluon propagator at
finite temperature with improved action. The preliminary results on the
small lattice $8^3 \times 4$ 
shows 
that in the confined phase, gauge copies are found in lattice version 
of Lorentz gauge and the gluon propagators show the peculiar behavior, which
Nakamura group\cite{Nakamura} has found on the large lattice with standard
action. \\
\noindent ACKNOWLEDGEMENT  \\                                      
This work is supported by the Supercomputer Project (No.97-27)
of High Energy Accelerator Research Organization (KEK).       
 We would like to express our                                   
thanks to the members of KEK for their warm hospitality.


\begin{thebibliography}{9}
\bibitem{Zubarev} D.N.Zubarev, Nonequilibrium Statistical \hspace *{\fill}\\ 
Thermodynamics Plenum, New York 1974.                  
\bibitem{Hosoya} A.Hosoya, M.Sakagami and M Takao,                 
                 Annals of Phys.154(1984) 229.                     
\bibitem{Karsch} F.Karsch and H.W.Wyld,Phys.Rev.\\
                D35(1987) 2518.  
\bibitem{Horsley} R.Horsley and W.Schoenmaker, Nucl. Phys. B280[FS18](1987),716, ibid.,735.    
%\bibitem{Masuda} N.Masuda, A.Nakamura, S.Sakai and F.Shoji,
%                 Nucl. Phys.B(Proc Suppl) 42(1995),526. 
\bibitem{Iwasaki} Y. Iwasaki preprint UTHEP-118,\\                 
                  S.Itoh, Y.Iwasaki, Y.Oyanagi and T. Yoshi\'e,    
                  Nucl. Phys. B274(1986), 33.\\                  
                  Y.Iwasaki, K.Kanaya, S.Sakai and T. Yoshi\'e,    
                  Nucl. Phys. B(Proc. Suppl.) 42(1995), 502.      
\bibitem{Sakai} A.Nakamura, S.Sakai and K. Amemiya,
                Nucl. Phys. B(Proc. Suppl) 53(1996),432
\bibitem{Gavin} S. Gavin, Nuclear Phys.A345(1985) 826.  
\bibitem{Kaneko} Y. Iwasaki, K. Kaneko, K. Kanaya and T. Yoshie
                Nucl. Phys.B(Proc. Suppl.) 53(1996) 429.  
\bibitem{Nakamura} A. Nakamura, H. Aiso, M. Fukuda, T. Iwatani
          T. Nakamura and M. Yoshida, 
          Proceedings of Confinement95(1996) 90.
\end{thebibliography}
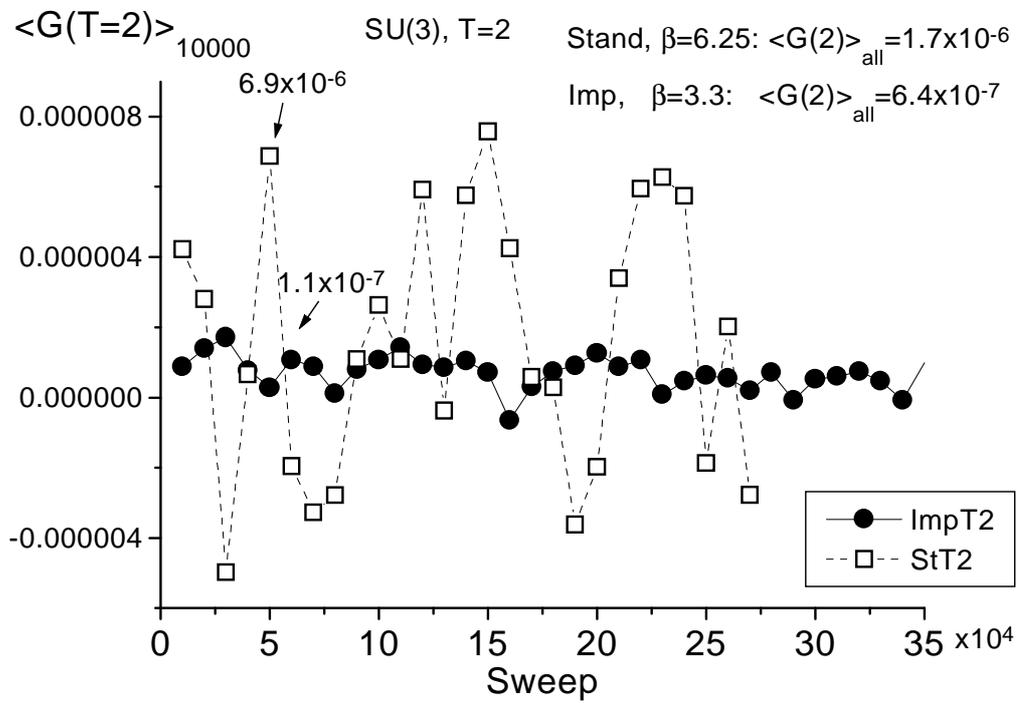
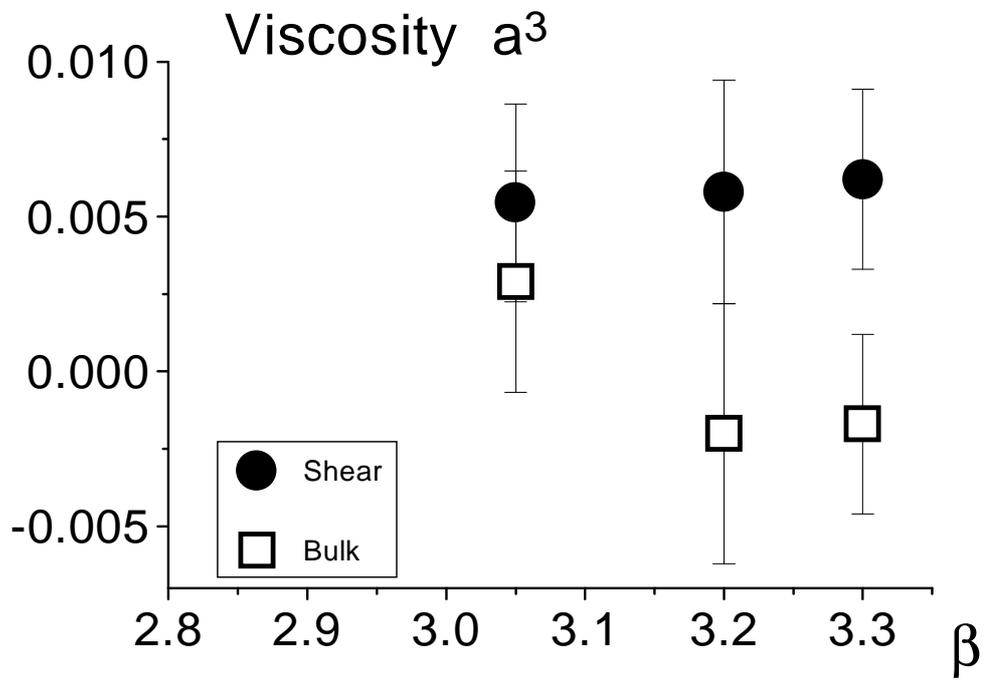
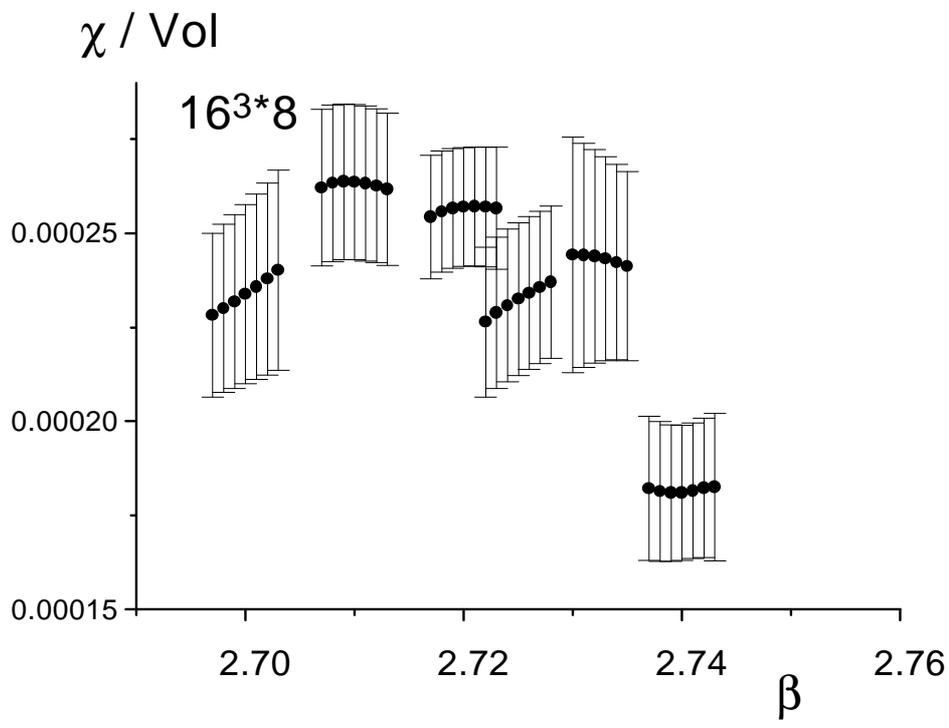
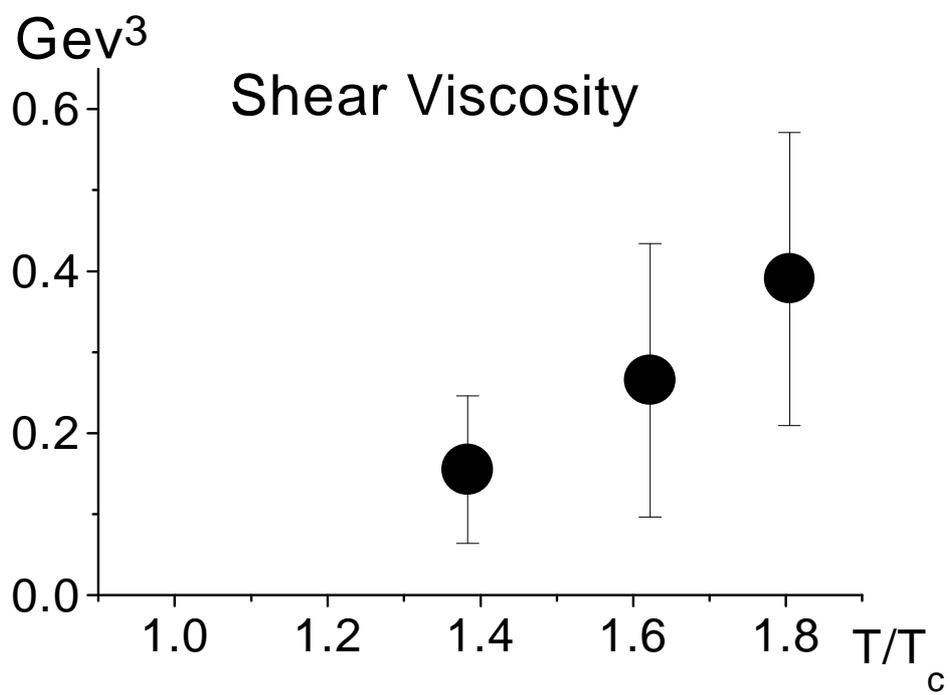
\end{document}